\documentclass[twocolumn,showpacs,preprintnumbers,amsmath,amssymb]{revtex4}

\usepackage{isomath}

\usepackage{upgreek} 

\usepackage{epsfig, amsmath,amsfonts, amssymb,graphicx,amsthm,color}

\usepackage{mathtools}
\usepackage{dcolumn}
\usepackage{bm}
\usepackage{rotating}
\usepackage{amssymb}
\usepackage[latin1]{inputenc}
\usepackage{graphicx}

\usepackage{epstopdf}
\begin{document}

\title{Bichromatic magneto-optical trapping for $ J\rightarrow J,J-1 $ configurations}
\author{Anne Cournol\footnote{Corresponding author: anne.cournol@u-psud.fr}, Pierre Pillet, Hans Lignier and Daniel Comparat}
\affiliation{Laboratoire Aim\'{e} Cotton, CNRS, Univ. Paris-Sud, ENS Cachan, Universit\'e Paris-Saclay, B\^{a}t. 505, 91405 Orsay, France }

\date{\today}

\begin{abstract}
A magneto-optical trap (MOT) of atoms or molecules is studied when two lasers of different detunings and polarization are used. Especially for  $J\rightarrow J,J-1$ transitions, a scheme using more than one frequency per transition and different polarization is required to create a significant force. Calculations have been performed with the simplest forms of the $J\rightarrow J-1$ case (i.e. $J''=1 \rightarrow J'=0$) and  $J\rightarrow J$ case (i.e. $J''=1/2 \rightarrow J'=1/2$). A  one dimensional (1D) model is presented and a complete 3D simulation using rate equations confirm the results. Even in the absence of Zeeman effect in the excited state, where no force is expected in the single laser field configuration, we show that efficient cooling and trapping forces are restored in our configuration. We study this mechanism for the C$_2^-$ molecular anion as a typical example of the interplay between the two simple transitions $J \rightarrow J,J-1$.
\end{abstract}

\maketitle

Laser cooling and magneto-optical trapping of atoms typically involves $J\rightarrow J+1$ closed transitions driven by counter-propagating circularly polarized laser fields ($J$ is the total angular momentum).
On the contrary laser forces on molecule, a more recent subject
despite the pioneer experiments \cite{1994ZhPmR..59..381V,1997PhRvL..79.2787S},  
 typically involve $N''=1 \rightarrow N'=0$ transitions  ($N$ is the rotational quantum number) between 
 $  X^2 \Upsigma (v''=0)$ and $ A^2\Uppi_{1/2} (v'=0) $ vibronic levels
 \cite{2013PhRvL.110n3001H,2014PhRvA..89e3416Z,2014Natur.512..286B}.
Including the electron spin leads to $J''=1/2,3/2 \rightarrow J'=1/2$ transitions.
 Theoretical study of the correct choice of circular polarization for magneto-optical trapping depending on the
angular momenta, $J''$ and $J'$, and on the $g$-factors, $g''$ and $g'$ (respectively of the lower and upper states) has been performed in Ref \cite{2015NJPh...17a5007T}. It has been found heuristically that the
trapping force is  weak whenever $g'$ is small compared to $g''$.
This is a serious limitation because in pure Hund's case (a) or (b), $A^2\Uppi_{1/2}$ does not present any Zeeman effect \cite{van2012manipulation}, so $g'=0$ prevents any force. However experiments, such as that on SrF \cite{2015NJPh...17c5014M}, were possible because of rotational and
spin-orbit  $\Uppi - \Upsigma$ 
mixing allowing a small  non zero value for $g'$ ($\sim 0.1$ \cite{2015NJPh...17a5007T}).
In addition to this weak force, $J''=1/2,3/2 \rightarrow J'=1/2$ transitions present one or two dark states 
 \cite{2002PhRvA..65c3413B,2004PhRvA..69c3410T} which leads to experimental difficulties if a stronger
confining force is desired. Yet stronger force can be produced by rapidly switching the magnetic field gradient and laser beam polarization on a timescale (typically in the sub-microsecond range which is not a simple experimental task) preventing the adiabatic following of the atomic states as done in reference \cite{2013PhRvL.110n3001H} . 

In this letter we suggest that efficient magneto-optical forces can be restored in all $J'' \rightarrow J'$ cases, even for $g'=0$, in a very simple way, by simply adding one laser field of opposite polarization with a different detuning. We treat
the simplest  $J\rightarrow J-1$ case (that is $J''=1 \rightarrow J'=0$) and  $J\rightarrow J$ case (that is $J''=1/2 \rightarrow J'=1/2$).
An analytical one dimensional (1D) model is presented and help get the physical insight of the process. A complete 3D simulation using rate equations confirms the results.
Although we do not consider any hyperfine structure, the principle should be very similar if $J$ is replaced by $F$, the total angular momentum including the nuclear spin.
A first study on the optical pumping using the simple $J''=1 \rightarrow J'=0$ configuration shows that the only possibility to produce a force in stationary regime requires non monochromatic laser fields with different polarization. We then show that a simple bichromatic laser scheme establishes significant trapping and cooling forces for two angular configurations,i.e. $J''=1 \rightarrow J'=0$  and $J''=1/2 \rightarrow J'=1/2$. Finally different schemes of magneto-optical trapping are investigated for the C$_2^-$ species, a typical example with both $J \rightarrow J$ and $J\rightarrow J-1$ transitions that cannot be cooled by a standard MOT configuration. Laser cooling and trapping of C$_2^-$ would be of high interest for the study of cold molecules, molecular anions, charged particle sources, antimatter physics as suggested in \cite{2015PhRvL.114u3001Y}.

Real MOT behavior can be very complex due to coherent population trapping into dark states, polarization gradients created by interference between laser beams or hyperfine state mixing by magnetic field. 
 We will not address these aspects, firstly because we do not consider dipolar forces or Sisyphean frictions
 \cite{prudnikov2004kinetics}, and secondly because several studies on standing wave lattices have shown that
  coherent (stimulated laser cooling or coherent population trapping) forces can indeed lead to "rectified" forces and provide confinement  for sufficiently high laser intensity  \cite{1987JETPL..46..332K,1990OptCo..77...19C,1990PhRvL..65.1415G,1991OptL...16.1695P,1992EL.....20..687E,1992PhRvL..68.3148H,1995OptCo.118..261P,1995PhRvA..51.1407D,1997PhRvL..78.1420S,2010arXiv1009.3118K}.  Similarly, we only study Doppler cooling even if laser fields considered in this letter can lead to sub-Doppler, velocity-selective coherent population trapping (VSCPT) or grey molasse  cooling (see Refs \cite{1992EL.....17..133V,1997PhRvA..56.3083R,2013PhRvA..87f3411G,2014PhRvA..90d3408B} and reference therein).

Our study is simpler and based on rate equations where none of these effects is included. This approach has been proven to well describe Doppler cooling of atoms. Hence, even though coherence effects may change some of the quantitative results obtained here, they are unlikely to change the main conclusions. The rate equations for the populations $\rho$ and the scattering force $\bm{F}$  used in our simulation  depend on the decay rate $\Gamma_{ i j}$ between an upper level $i=|J',M'=M_i \rangle$ and a lower one $j=|J'',M''=M_j\rangle$:
\begin{align}
  	&\dot \rho_{i}  =   \sum_{j} \left[  
			-  \Gamma_{i j}  \rho_{i}  +  \gamma_{ i j} (\rho_{j} -\rho_{i} ) 
		\right] \label{excited_pop} \\
		&\dot	 \rho_{\mathrm j }  =   \sum_{i} \left[ \Gamma_{i j}  \rho_{i} +  \gamma_{i j} (\rho_{i} -\rho_{j} ) 
		\right] \label{ground_pop}\\
		&	\bm{F}=  \sum_{\mathrm L}  \hbar  \bm k_{\mathrm L}  \sum_{i,j}  \gamma_{i j}^{\mathrm L}  (\rho_{j}  -   \rho_{i} )  \label{force_vs_pop}
\end{align}
where $\gamma_{i j}=\sum_{\mathrm L}  \gamma_{i j}^{\mathrm L} $ and
 $  \gamma_{i j}^{\mathrm L}  =  \frac{ \Gamma}{2}
 \frac{2 | \Omega_{i j}^{\mathrm L} |^2/\Gamma^2}{1+4 {\delta_{i j}^{\mathrm L}}^2/\Gamma^2}
 $ is the rate of excitation and stimulated emission of the transition between states $i$ and $j$ due to the laser L 
  with a linewidth $\Gamma_{\mathrm L}\ll \Gamma$. 
 The Rabi frequency $ \Omega_{i j}^{\mathrm L} $ depends on the electric dipole moment $\mu_{i j}  $   between the states and the electric field amplitude $E^{\mathrm L}_{p=M_j - M_i}$  expressed in the helicity basis $\bm{e}_p$.  
	The detuning for the laser $\mathrm L$ is given by $\delta_{\mathrm i j}^{\mathrm L} = \delta_0 - {\bm k_{\mathrm L}} \cdot {\bm v} +  \mu_{\mathrm B} B (g' M' - g'' M'')/\hbar$.
The laser saturation is defined by $s^{\mathrm L} = 2 | \Omega^{\mathrm L} |^2/\Gamma^2 $ and the fractional strength of the transition being 
	$ f_{i j} = {\Omega_{i j}^{\mathrm L}}^2/{\Omega^{\mathrm L}}^2 = \mu_{i j}^2/ \mu^2 =  \Gamma_{i j}/\Gamma = |\langle J'' M'', 1 p | J' M' \rangle|^2$.
	A given laser irradiance $I$ leads to $s = I/I_{\mathrm{s}}$ where $I_{\mathrm{s}} = \pi h c \Gamma/(3 \lambda^3)$ 
 is the saturation intensity for a transition of wavelength $\lambda$.

 Two simulations are performed by solving exactly the previous equations: using either an analytical approach for the 1D case or a Kinetic Monte Carlo method described in Ref. \cite{2014PhRvA..89d3410C} for the 3D case.
 
  In the 1D case, the laser field can be separated into two beams, L$^+$ and L$^-$, travelling in opposite directions. The physical insight of this configuration is easily obtained from the steady-state solution at low intensity  ($s^{\mathrm L} \ll 1)$) regime, where the scattering force becomes:
 \begin{equation}
 \bm{F}  \approx  \hbar  \bm{k}  \sum_{j}  \frac{\gamma_{j}^{+} -\gamma_{j}^{-}}{\gamma_{j}^{+} +\gamma_{j}^{-}}  \sum_{i} \Gamma_{i j} \rho_{i}   \label{eq_force}
 \end{equation}
	with  $\gamma_{\mathrm  j}^\pm =\sum_{i} \gamma_{i j}^{\mathrm{L^\pm}}$.

\begin{figure*}
\mbox{
\centering

\includegraphics*[width=0.80\textwidth]{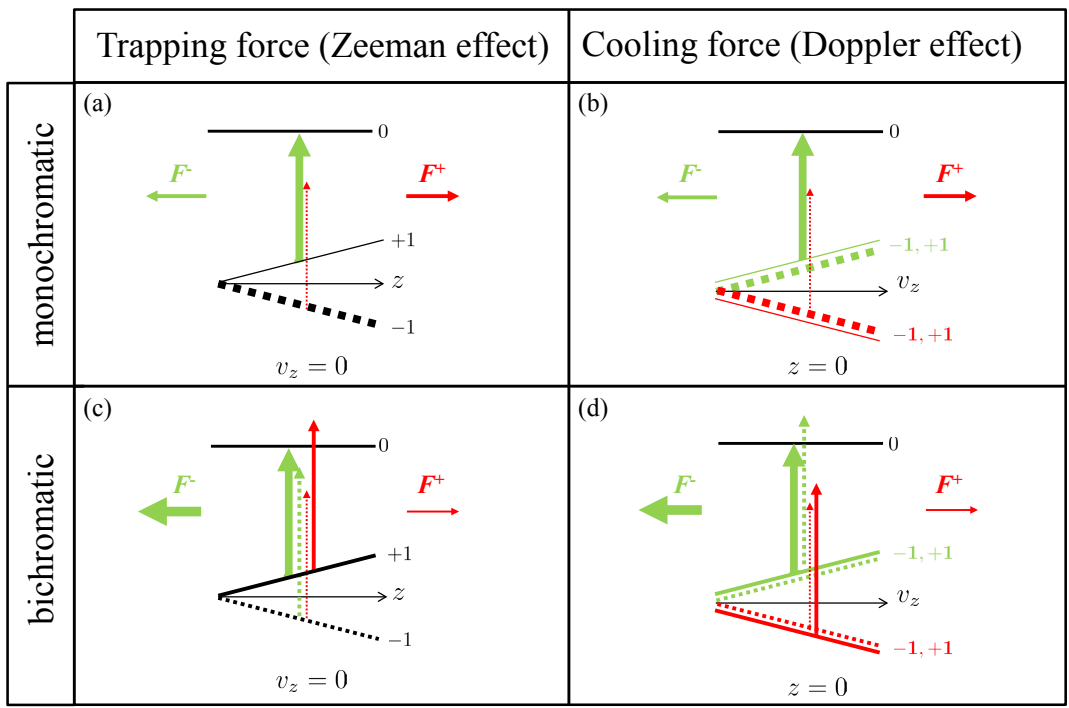}
	}			
		\caption{(Color online). One dimensional laser cooling in the $J''=1 \rightarrow J'=0$ configuration ($g''<0$). 
Green (light gray) and red (dark gray) respectively refer to lasers propagating to the left ($L^-$) and right ($L^+$). Dotted lines are used for $M''=-1$ levels and the corresponding driving transition $\sigma^+$ whereas solid lines represent $M''=+1$ levels and transitions $\sigma^-$. The total force $\bm{\mathrm{F}}=\bm{\mathrm{F^+}}+\bm{\mathrm{F^-}}$, resulting from lasers coming from $z>0$ and $z<0$, is given by Eq. \ref{force_vs_pop} which combines the coupling strengths $\gamma^{\pm}_{\pm 1}$ (schematically indicated by the thickness of the vertical arrows) and the amounts of population $\rho_{\pm 1}$ in the $M''=\pm1$ levels (schematically indicated by the thickness of lines representing the internal states). Panels (a) and (b): standard case with one $\sigma^+ - \sigma^-$ red-detuned laser field. Panels (c) and (d): an extra  $\sigma^- - \sigma^+$ laser field on resonance (for $z=v=0$) is added to the standard case. For trapping (in $v=0$), see panels (a) and (c): the energy of $M''=\pm 1$  depends on the position because of the Zeeman effect. For cooling (in $z=0$), see panels (b) and (d) where the Doppler effect shifts the two $M''=\pm 1$ states in the same way. Because of the two directions of propagation, there are two sets (red and green) of shifted states. 
		}
		
\label{fig1}
\end{figure*}

However, even for low velocities and monochromatic fields, a general theoretical description of radiative forces is very complex \cite{1991PhRvA..44..462N,1998EL.....44...20N,1998OptCo.148..151B,2004PhRvA..69c3410T}. We thus study the simplest relevant case: the  1D  $J''=1 \rightarrow J'=0$ transition. In this model, no laser polarization can induce $\pi$ transitions: thus $M''=0$  is not coupled to the upper state and can be excluded in first place leading to a simple $\mathrm{\Lambda}$ three level structure (see Fig. \ref{fig1}).  It is important to stress that spontaneous emission would accumulate population in $M''=0$ and no force would be produced in a real 1D experiment. Thus techniques using destabilization of dark states with off axis magnetic field or lasers should be used \cite{2002PhRvA..65c3413B}. The $\mathrm{\Lambda}$ case has a single upper state $i=0=|J'=0,M'=0 \rangle$ decaying toward the two lower levels $j=\pm 1=|J''=1,M''=\pm 1\rangle$ with a rate $\Gamma/2$.
The laser configuration must produce a friction force bringing the particle velocity to zero and a trapping force bringing its position back to zero. Whatever the physical reasoning is for $z>0$ and $v>0$ (where necessarily the needed force is dominantly produced by L$^-$), it is also valid for $z<0$ and $v<0$ by reversing the roles of L$^-$ and L$^+$. Thus, for the sake of symmetry, each pair of left L$^+$ and right L$^-$ laser beams will have the same power and detuning but opposite polarizations.

Let us first look at the configuration using two counter-propagating red-detuned laser beams with $\sigma^+$ and $\sigma^-$ polarization. Figs. \ref{fig1}(a),(b) show that, for any position (resp. velocity), one of the laser frequencies is closer to resonance than the other because of the Zeeman effect (resp. Doppler effect). Through optical pumping, there is an accumulation of population in the sub-level further from resonance and the excitation rate is weaker for this sub-level than for the less populated one. The imbalance in populations and in excitation rates are exactly compensated, which consequently equals the scattering rates from the restoring and anti-restoring beams: the resulting (trapping or friction) force is zero \cite{Minogin1985,1991OptL...16.1695P}. This conclusion is valid at any position and velocity whatever the optical detuning and the magnetic field gradient (see Fig. \ref{fig1}) as it can be confirmed by using Eq. (\ref{eq_force}): a given level $M''=\pm 1$ is coupled with a single polarization to the upper state $M'=0$ and is thus driven solely by lasers from a single side, so $ \sum_{j} \frac{\gamma_{j}^{+} -\gamma_{j}^{-}}{\gamma_{j}^{+} +\gamma_{j}^{-}} = (1-1)=0$. Note that even with polychromatic fields and unique polarization for any direction, the equations are exactly the same and consequently the total force vanishes.

The use of nonpure polarized light might restore a force \cite{prudnikov2004kinetics}. However, in such a case, only a friction force can be recovered. Effectively the trapping force would remain null, as can be understood  by using Fig. \ref{fig1}(c), if the extra laser with complementary polarization had the same detuning as laser of Fig. \ref{fig1}(a)  (see also the $\delta=-\Gamma$ case of Fig. \ref{fig:parameters}). Thus, the only remaining possibility is to use polychromatic laser fields with different polarization. This approach is closely related to studies of  magneto-optical trap (mainly on Na or Rb  \cite{1993PhRvA..47.4563M,1995OptL...20.2529F,1997JaJAP..36.5310Z,1997PhRvA..55.4621O,2001PhRvA..64b3412N,2007JaJAP..46L.492T,2008PhRvA..78f3421T}) with cooling and repumping lasers either tuned on (i) pure D$_2$ (standard type I MOT using $F\rightarrow F+1$, or more interestingly for our purpose, type II MOT using $F\rightarrow F-1$ or $F\rightarrow F$ transitions \cite{prentiss1988atomic}), (ii) pure $D_1$ \cite{1996JPhB...29.3051M,1997OptCo.135..269F} or (iii) D$_1$ and D$_2$ (type III MOT \cite{1999PhRvA..59.3101M,2001EPJD...13...71A}) \footnote{Not to be confused with Type I-II and III used in non alkali atoms and using narrow lines and metastable state \cite{2002PhRvA..66a3411L}}. However, these studies only consider the case where the two laser fields drive different levels. Here we treat the case where both laser beams, with different polarizations, drive the same level in a so-called  dual (or bichromatic or two color) magneto-optical trap. 
In a recent study of magneto-optical trapping of CaF molecules published in \cite{2015PhRvA..92e3401T}, it has been found that such a configuration should allow one to realize a MOT of CaF. If the authors put forth the proposition to benefit from this effect to enhance MOT forces, we aim at giving requirements and details to get such forces as well as the underlying mechanism for both $J \rightarrow J$ and $J \rightarrow J-1$ schemes.

This dual frequency and polarization is in fact a natural choice to create a force. Let us assume that all the population is in $M''=+1$ at $z>0$. A restoring force occurs because L$^-$ is absorbed but, as explained before, optical pumping  to $M''=-1$ quickly stops the force. However, the force can be easily restored before the end of the optical pumping, by a rapid switching of L$^-$ detuning and polarization, such as L$^-$ becomes more resonant with the now populated $M''=-1$.  L$^-$ is thus always  dominant and the particles are pushed toward the center. Repeating the switching process (obviously, due to symmetry, the same must be done with L$^+$) restores  the force dynamically. This simple and efficient way of solving the problem is similar to switching both the magnetic field gradient and the laser beam polarization as done in Ref. \cite{2013PhRvL.110n3001H}, but is easier to implement experimentally. This approach has been tested with our Monte Carlo code; in the case considered here, we found that the forces are very similar to those produced when both lasers are present at the same time because a quasi-stationary regime is often reached. Therefore in the following we only consider continuous and stationary laser configurations.

\begin{figure}
\mbox{
\centering

\includegraphics*[width=\columnwidth]{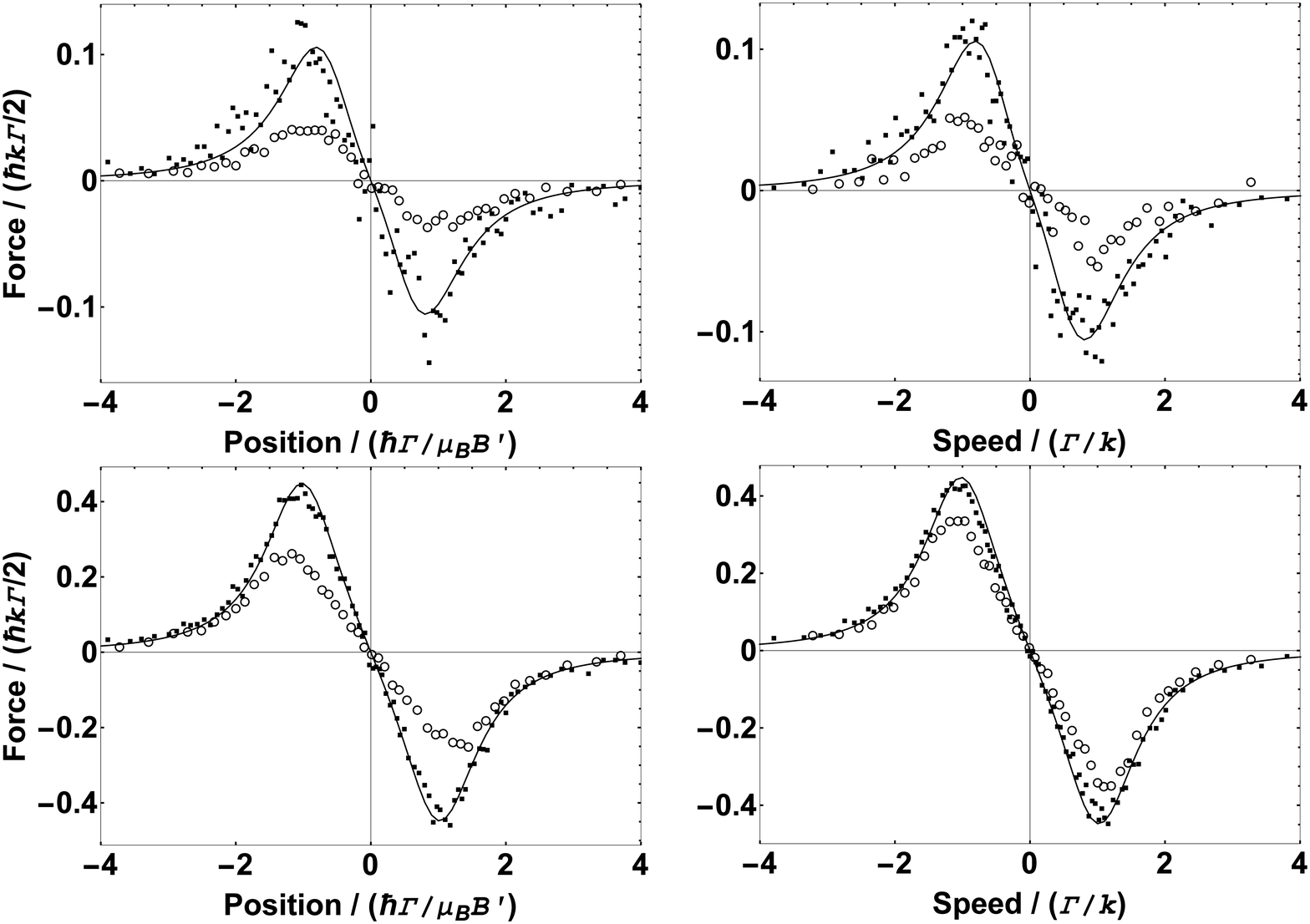}
	}
		\caption{Upper Panels: trapping and cooling forces produced by the bichromatic configuration with $\sigma^+$, $\sigma^-$ polarization and $g'' =-1$. The results correspond to the case of Fig. \ref{fig1} (c,d). Lower panels: for comparison, forces obtained with the standard type I MOT $J''=0 \rightarrow J'=1$ with $\sigma^+-\sigma^-$ single laser beam configuration and $g'=-1$. Solid lines: analytical 1D simulation. Squared dots: Monte Carlo 1D simulation. Open circle: Monte Carlo 3D simulation. Simulations are performed with saturation parameters $s=1$ detunings $\delta = -\Gamma$ and $\delta=0$ when a second laser is present.
}
\label{fig:forces}
\end{figure}

One  of the simplest configuration producing trapping and friction force is shown in Fig. \ref{fig1} (c),(d)  where two L$^-$ lasers are (i) a $\sigma^-$, red-detuned  laser and (ii) a $\sigma^+$ laser on resonance (at $z=v=0$). We have simulated this situation by solving the stationary rate equations analytically and results are presented in Fig. \ref{fig:forces} with comparison to the standard type I MOT ($J''=0 \rightarrow J'=1$, $\sigma^+-\sigma^-$ single laser beam configuration). We also plot results from 1D and 3D Monte Carlo simulations where the forces (acceleration) are calculated for a given initial position or velocity after an evolution time which is short enough to limit the change of initial positions or velocities but long enough to extract accelerations. In all cases, the trapping and friction forces  are strong and only slightly smaller than the standard type I MOT. The physical origin of the force is provided by lower panels in Fig. \ref{fig1}: the restoring force is effective because the L$^-$ lasers, that bring populations back to $z=v=0$, always have a component that is more resonant than the anti-restoring lasers (L$^+$). The 3D simulation shows that adding the $\pi$ polarization coming from other beams that repump the $M''=0$ population, does not significantly modify the physical comprehension of the process but simply reduces the force by a factor $2-3$ since the absorption and emission processes are shared between the three directions.

 \begin{figure}
\mbox{
\centering

\includegraphics*[width=\columnwidth]{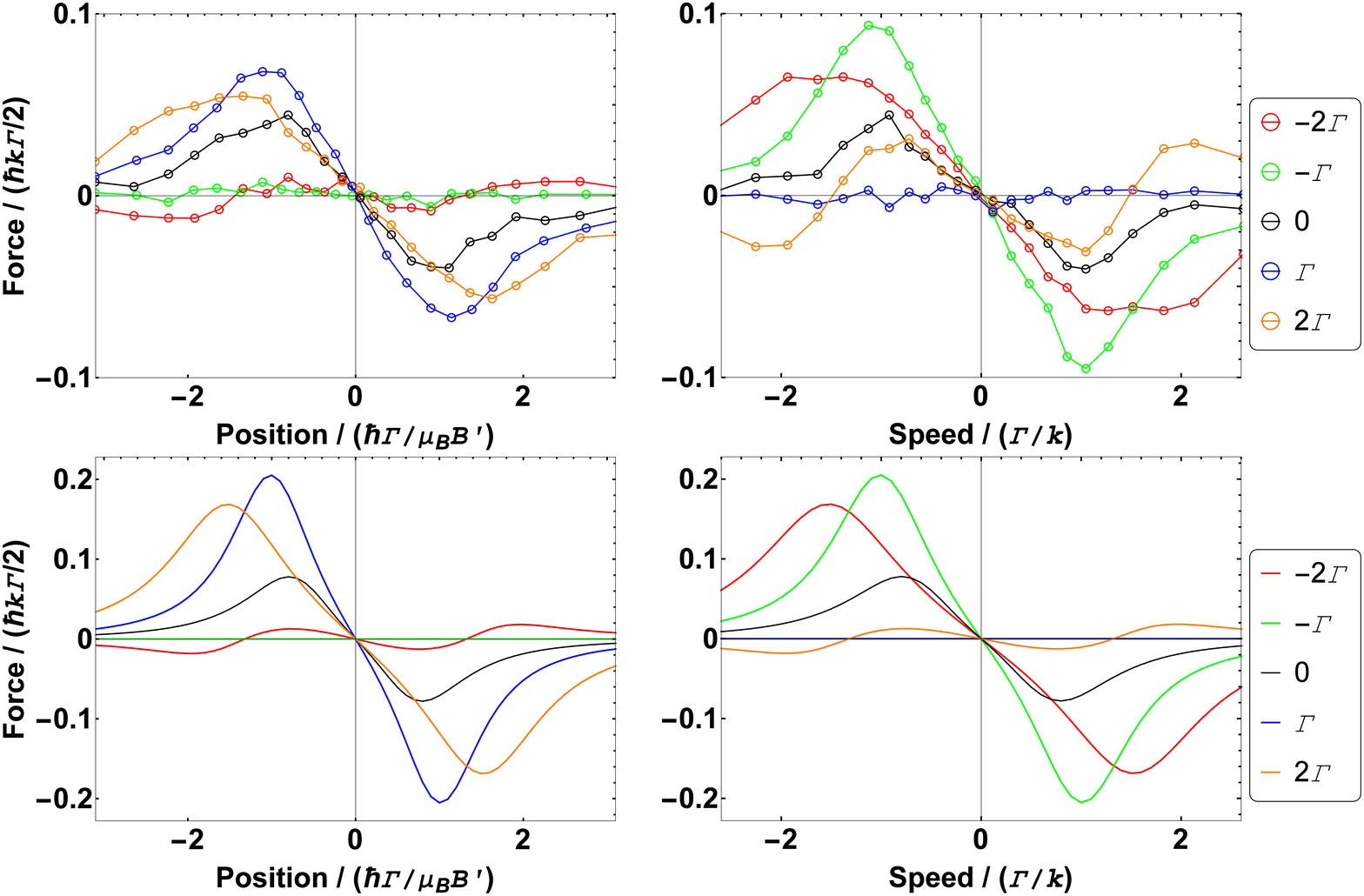}
	}	
		\caption{(Color online). Dependence of the trapping (left) and cooling force (right) on different values of the second laser detuning.	Parameters are similar to those used in Fig. \ref{fig:forces} (upper panel 3D simulation, lower 1D analytical solutions).
}
\label{fig:parameters}
\end{figure}

The dependence of the forces on the power of the second (frequency component) laser is rather intuitive (it must have a decent power compared to the first laser) whereas the dependence on the detuning value is less obvious. Therefore we plot the forces in the 1D and 3D cases for several detuning values of  the second laser in Fig. \ref{fig:parameters}.  The detuning of the second laser is an important parameter but it acts on the trapping or cooling forces differently.

 \begin{figure}
\mbox{
\centering
\includegraphics*[width=0.95\columnwidth]{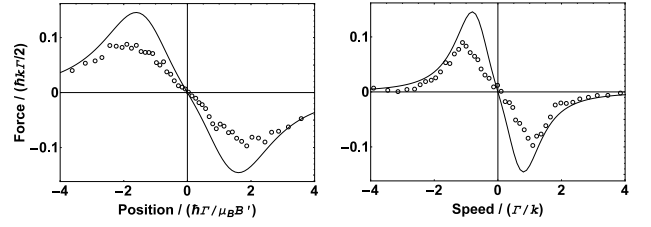}
	}	
			\caption{Trapping and velocity forces with dual laser frequency and polarization on the $J''=1/2 \rightarrow J'=1/2$ configuration with $g' = 0$ and $g''=1$. The other parameters are the same as those used in Fig. \ref{fig:forces}. The 1D case (solid lines) is solved analytically while the 3D case (open circles) results from the Monte-Carlo simulation. }
\label{fig:J1/2}
\end{figure}

For the $1 \rightarrow 0$ transition, Fig.  \ref{fig1} gives  an important hint to understand the physical nature of the forces. Indeed, for trapping, our discussion about the optical pumping already indicates that a second L$^-$ laser, $\sigma^+$ polarized, (implying the existence of its  symmetric L$^+$, $\sigma^-$) is required to "repump" the population into $M=+1$. In Fig. \ref{fig1} (c), the Zeeman effect adds a detuning $\pm \Gamma$ depending on the laser polarization $\sigma^{\pm}$. A "naive" idea would be to shift the frequency of this supplementary laser by $+\Gamma$ in order to make it resonant. This detuning would lead to a maximum trapping force (see the blue curve corresponding to $\delta=+\Gamma$ in the upper right panel, Fig. \ref{fig:parameters}). However,  for cooling (see Fig. \ref{fig1} (d)) the situation is different because the Doppler effect adds a detuning $\pm \Gamma$ depending on the laser axis L$\pm$. Thus, this second L$^-$ laser would be too far from resonance ($+2 \Gamma$) whereas its symmetric L$^+$, $\sigma^-$ would be tuned to resonance, thus compensating the effect of the first L$^-$ laser and resulting in the absence of cooling force (as shown in Fig. \ref{fig:parameters}). In summary, a red detuning favors a cooling force and a blue detuning favors a trapping force. The choice of a resonant laser is a reasonable compromise and has been used in Fig. \ref{fig1} and \ref{fig:forces}. Even in this simple case, the force dependence on the position and velocity is complicated. This indicates that, in general, optimization of the laser detuning and polarization is not straightforward: a stronger force can stand alongside with a smaller capture range. It is thus important to avoid any perturbative approach in position and velocity.

The physical mechanism behind a bichromatic MOT is very general. Obviously it can be used for all $J \rightarrow J-1$ schemes, such as $3/2\rightarrow 1/2$, but it is also interesting to show that it could be useful for $J \rightarrow J$ schemes.
In order to quickly illustrate the $J \rightarrow J$ case, we have chosen the simplest one, namely $J=1/2 \rightarrow J=1/2$.
Here again, no force appears in the single frequency pure polarization case \cite{prudnikov2004kinetics} but a bichromatic MOT produces quite substantial forces that are plotted in Fig. \ref{fig:J1/2} \footnote{Because  $g' M' - g'' M'' = \mp (g'+g'')$ for a  $\sigma^\pm$  polarization, the forces scale with $g'+g''$.}.

\begin{figure*}
\mbox{
\centering
\includegraphics*[width=1.5\columnwidth]{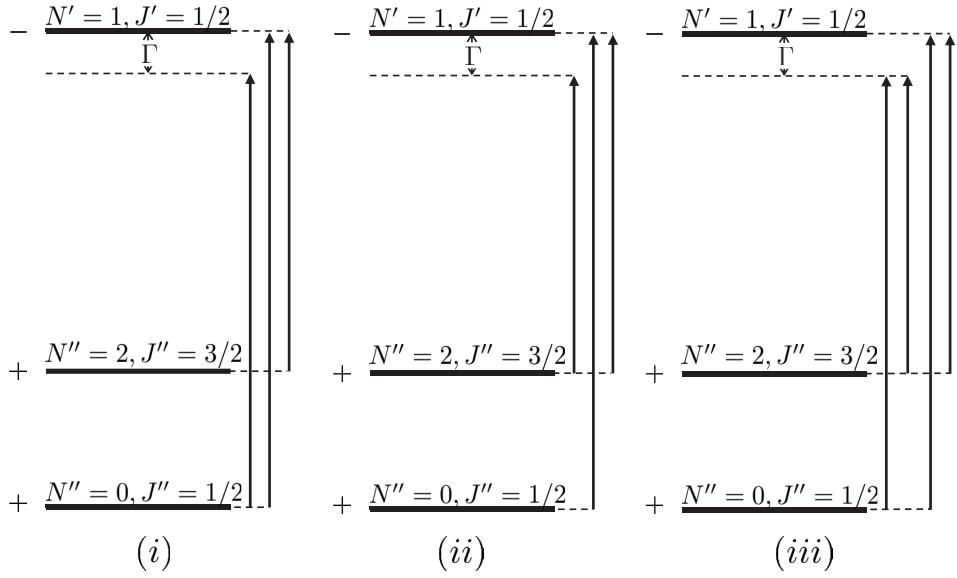}
	}	
			\caption{Three different bichromatic MOT schemes for C$_2^-$ using red-detuned cooling lasers driving the transitions (i) $J''=1/2 \rightarrow J'=1/2$, (ii) $J''=3/2 \rightarrow J'=1/2$ and (iii) their combination. Symbols $J$, $N$ and $\pm$ respectively denote the total angular momentum, the rotational quantum number, and the total parity of the molecular wave-function. For each scheme, the two lower levels belong to X$^2 \Upsigma(v'')$ while the upper level belongs to B$^2\Upsigma (v'=0)$.}
\label{fig:C2minus_schemes}
\end{figure*}

\begin{figure*}
\mbox{
\centering
\includegraphics*[width=2\columnwidth]{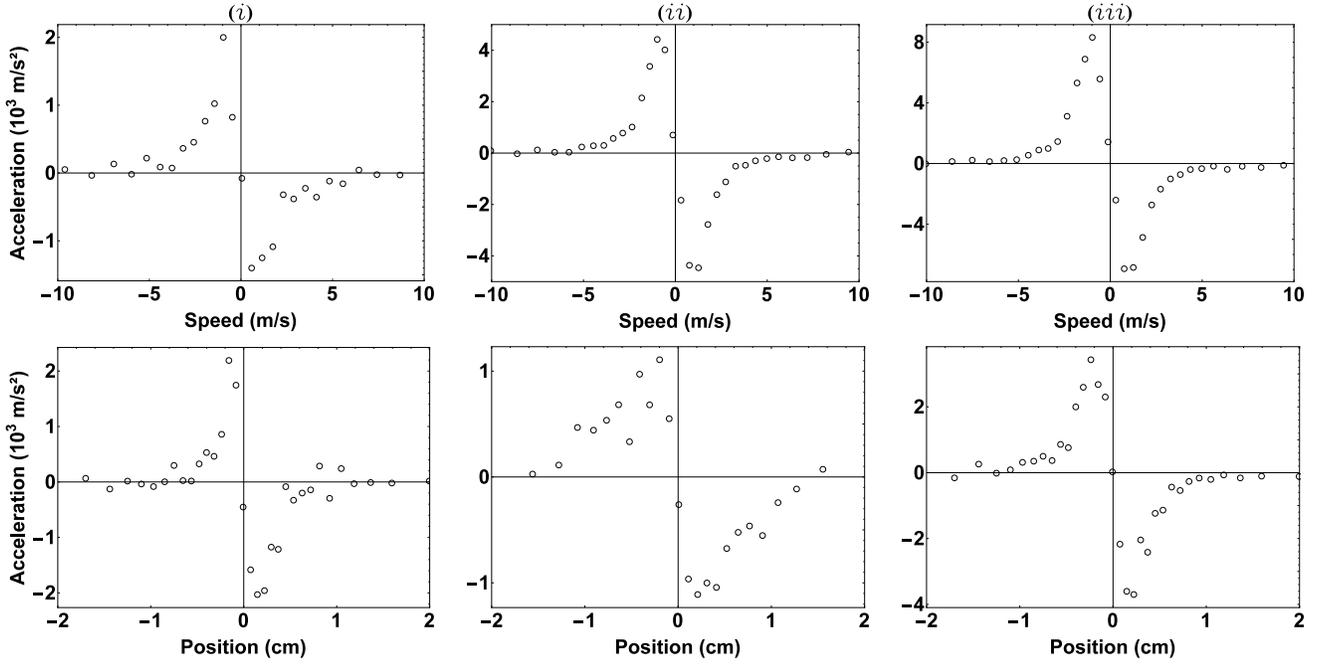}
	}	
			\caption{Acceleration as a function of position and speed (in SI units) for three different bichromatic configurations using C$_2^-$ parameters. These configurations correspond to the standard MOT lasers, producing no force alone, and lasers tuned on (i) $J''=1/2 \rightarrow J'=1/2$ (left), (ii) $J''=3/2 \rightarrow J'=1/2$ (center) and (iii) their combination (right).}
\label{fig:C2minus}
\end{figure*}

Finally, attention is turned to C$_2^-$ molecular anion for which the most abundant isotopologue, i.e. $^{12}$C$_2^-$, does not have any hyperfine structure. In its B$^2\Upsigma (v'=0)$ $\leftrightarrow$ X$^2 \Upsigma(v'')$ vibronic transition, this anion exhibits a combination of $J\rightarrow J-1$ and $J\rightarrow J$ schemes that constitutes a closed rotational system if the driven transitions couple $J''=1/2,\ 3/2$ of the $N''=0,2$ ground states to  $J'=1/2$ of the $N'=1$ excited state as shown in Fig. \ref{fig:C2minus_schemes}. Effectively, with such an exciting scheme, the rovibronic transition rules ensure that the excited state decays, by spontaneous emission, to $J''=1/2,\ 3/2$ of the ground state. Magneto-optical trapping this species in single frequency MOT case is not expected because of the quadratic dependence of the Zeeman effect in the excited state, i.e. $g' =0$ at first order. This is the main reason why authors of Ref. \cite{2015PhRvL.114u3001Y} do not attempt to simulate trapping and only focus on molasse cooling. Details of practical realization of C$_2^-$ cooling have been discussed in \cite{2015PhRvL.114u3001Y}. Here we show that the trapping problem can be solved by a bichromatic MOT. In our study, the physical picture  is restricted to one dimension and to the X$^2 \Upsigma (v''=0)$ $\leftrightarrow$ B$^2\Upsigma (v'=0)$ transition at 541 nm (adding vibrational repumping lasers would not change the physical mechanisms because they drive the same excited state) with an excited state lifetime of 75 ns. Cooling and trapping C$_2^-$ is simulated with all the lasers at an intensity  of 1.8 mW/cm$^2$ with a uniform intensity for the transverse profile. For trapping, the magnetic field gradient is 10 G/cm.
Because the level structure of C$_2^-$ consists of the two systems previously studied ($J\rightarrow J$ and $J\rightarrow J-1$), it offers the possibility to address them individually or all together, which means that one or two red-detuned lasers are used for the cooling and friction forces (see Fig. \ref{fig:C2minus_schemes}). 
In any case, two lasers tuned to the resonant frequencies of the two $J''\rightarrow J'$ transitions at $z=0$ and $v=0$ must be added to give rise to the bichromatic MOT effect and avoid population trapping in of the $J''$ states.
In Fig. \ref{fig:C2minus}, forces resulting from schemes (i) and (ii) are presented. The friction force is larger for scheme (i) than for scheme (ii), which is due to a more favourable branching ratio for $J''=3/2 - J'=1/2$ than for $J''=1/2 - J'=1/2$. It is also noteworthy that the range over which the trapping force acts is different: it can be explained by  the Landé factor that is approximately four times smaller for $J''=1/2$ than for $J''=3/2$. The results obtained with the scheme (iii), shown in the right panels of \ref{fig:C2minus}, combine the best features of cases (i) and (ii): the trapping range is slightly increased by the $J''=3/2$ system and the amplitude of the force is as large as it may be with schemes (i) or (ii). In summary, optical pumping of a complex system allows one to create both cooling and trapping forces that were prevented in standard MOT configuration. This is not surprising because the key physical process of bichromatic MOT is to prevent accumulation of population in dark states by pumping populations of all Zeeman sub-levels optically. It is also interesting to note that combining several cooling lasers in presence of the necessary repumping lasers can improve the specifications of the trapping and cooling forces.

In conclusion, we have described a cooling mechanism based on two lasers with different types of polarization and frequencies driving the same transition scheme. This scheme produces friction and trapping forces for all possible transitions $J\rightarrow J-1 , J , J+1 $ even when the upper state does not present any Zeeman effect (yet, in this case, the lower level has to undergo some Zeeman effect). This cooling mechanism could be tested on $F=1\rightarrow F=0$ transition of $^{87}$Rb \footnote{Similar experiment has been performed on Na, but the $F=1$ upper level is too close to the $F=0$ to be neglected \cite{2001PhRvA..64b3412N}}. More interestingly, it could be used to cool and trap diatomic molecules on the $  X^2 \Upsigma (N''=1,v''=0) \rightarrow A^2\Uppi_{1/2} (N'=0,v'=0) $ transition, molecular negative ions such as C$_2^-$ studied in this article, and also some atoms, such as La$^-$ that exhibits a $J=2 \rightarrow J=1$ transition \cite{jordan2015high}. In the experimental configuration used for SrF \cite{2014Natur.512..286B,2015NJPh...17c5014M}, two lasers with opposite polarization have been used because of the opposite Land{\'e} factor; it seems probable that the process discussed here plays a significant role in the efficient trapping and cooling due to extra near resonance frequencies provided by sidebands originally created to cover the hyperfine splitting \cite{2015PhRvA..92e3401T}. Adding other lasers may even improve cooling and trapping, especially for higher $J$ values or more complicated schemes involving a hyperfine structure or non-linear Zeeman shift. Preliminary tests on a trichromatic scheme already indicate a significant enhancement of the forces.

We are indebted to L. Pruvost, P. Cheinet, W. Maineult and H. Lehec for useful discussions. 
The research leading to these results has received funding from the European
Research Council under the grant agreement n.~277762 COLDNANO, ANR MolSysCool and DIM Nano-K CMPV.

\bibliographystyle{h-physrev}

\bibliography{2015_bibli_sept_v2}

\end{document}